\documentclass[twoside,11pt]{article}

\usepackage[accepted]{melba}

\usepackage{rotating}
\usepackage[dvipsnames]{xcolor}

\usepackage{lscape}
\usepackage{amsmath,amsfonts}
\usepackage{multirow}
\usepackage{url}

\melbaid{2024:016}  
\doi{https://doi.org/10.59275/j.melba.2024-3d4e}
\melbaauthors{Hemforth et al.}  
\volume{2}
\firstpageno{1004}  
\melbayear{2024}  
\datesubmitted{02/2024}  
\datepublished{07/2024}  

\ShortHeadings{Automatic rating of incomplete hippocampal inversions (IHI)}{Hemforth et al.}

\title{Automatic rating of incomplete hippocampal inversions evaluated across multiple cohorts}

\author{{Lisa Hemforth,$^{*}$$^a$,
    Baptiste Couvy-Duchesne,$^a$$^c$,
    Kevin De Matos$^a$,
    Camille Brianceau$^a$,
    Matthieu Joulot$^a$, 
    Tobias Banaschewski$^d$, 
    Arun L.W. Bokde$^e$, 
    Sylvane Desrivières$^f$,
    Herta Flor$^g$,$^h$, 
    Antoine Grigis$^i$, 
    Hugh Garavan$^j$, 
    Penny Gowland$^k$, 
    Andreas Heinz$^l$,
    Rüdiger Brühl$^m$, 
    Jean-Luc Martinot$^n$, 
    Marie-Laure Paillère Martinot$^n$,$^o$, 
    Eric Artiges$^n$,$^p$,
    Dimitri Papadopoulos$^i$, 
    Herve Lemaitre$^i$,$^q$, 
    Tomas Paus$^r$,$^s$, 
    Luise Poustka$^t$, 
    Sarah Hohman$^aa$,
    Nathalie Holz$^d$, 
    Juliane H. Fröhner$^u$, 
    Michael N. Smolka$^u$, 
    Nilakshi Vaidya$^v$, 
    Henrik Walter$^l$,
    Robert Whelan$^w$, 
    Gunter Schumann$^v$,$^x$, 
    Christian Büchel$^y$, 
    JB Poline$^z$, 
    Bernd Itterman$^m$,
    Vincent Frouin$^i$, 
    Alexandre Martin$^a$, 
    IMAGEN study group, 
    Claire Cury$^b$,$^{**}$ and
    Olivier Colliot$^a$,$^{**}$}
    \\\\\\
    \scriptsize{
        $^{a}$Sorbonne Université, Institut du Cerveau – Paris Brain Institute - ICM, CNRS, Inria, Inserm, AP-HP, Hôpital de la Pitié-Salpêtrière, F-75013, Paris, France,
        $^{b}$University of Rennes, Inria, CNRS, Inserm, IRISA UMR 6074, Empenn ERL U-1228, Rennes, 35000, France
        $^{c}$Institute for Molecular Bioscience, the University of Queensland, Brisbane, 4072, Australia
        $^{d}$Department of Child and Adolescent Psychiatry and Psychotherapy, Central Institute of Mental Health, Medical Faculty Mannheim, Heidelberg University, Mannheim, 68159, Germany
        $^{e}$Discipline of Psychiatry, School of Medicine and Trinity College Institute of Neuroscience, Trinity College Dublin, Dublin, Ireland
        ${f}$Centre for Population Neuroscience and Precision Medicine (PONS), Institute of Psychiatry, Psychology $\&$ Neuroscience, SGDP Centre, King’s College London, London, United Kingdom
        $^{g}$Institute of Cognitive and Clinical Neuroscience, Central Institute of Mental Health, Medical Faculty Mannheim, Heidelberg University, Mannheim, 68159, Germany
        $^{h}$Department of Psychology, School of Social Sciences, University of Mannheim, Mannheim, 68131, Germany
        $^{i}$NeuroSpin, CEA, Université Paris-Saclay, Gif-sur-Yvette, France
        $^{j}$Departments of Psychiatry and Psychology, University of Vermont, Burlington, Vermont, 05405, USA
        $^{k}$Sir Peter Mansfield Imaging Centre School of Physics and Astronomy, University of Nottingham, University Park, Nottingham, United Kingdom
        $^{l}$Department of Psychiatry and Psychotherapy CCM, Charité – Universitätsmedizin Berlin, corporate member of Freie Universität Berlin, Humboldt-Universität zu Berlin, and Berlin Institute of Health, Berlin, Germany
        $^{m}$Physikalisch-Technische Bundesanstalt (PTB), Braunschweig and Berlin, Germany
        $^{n}$Institut National de la Santé et de la Recherche Médicale, INSERM U 1299 "Trajectoires développementales $\&$ psychiatrie", University Paris-Saclay, CNRS; Ecole Normale Supérieure Paris-Saclay, Centre Borelli, Gif-sur-Yvette, France
        $^{o}$Sorbonne University, Department of Child and Adolescent Psychiatry, Pitié-Salpêtrière Hospital, Paris, France
        $^{p}$Psychiatry Department, EPS Barthélémy Durand, Etampes, France
        $^{q}$Institut des Maladies Neurodégénératives, UMR 5293, CNRS, CEA, Université de Bordeaux, Bordeaux, France
        $^{r}$Departments of Psychiatry and Neuroscience, Faculty of Medicine and Centre Hosptalier Universitaire Sainte-Justine, University of Montreal, Montreal, Quebec, Canada
        $^{s}$Departments of Psychiatry and Psychology, University of Toronto, Toronto, Ontario, Canada
        $^{t}$Department of Child and Adolescent Psychiatry and Psychotherapy, University Medical Centre Göttingen, Göttingen, Germany
        $^{u}$Department of Psychiatry and Neuroimaging Center, Technische Universität Dresden, Dresden, Germany
        $^{v}$Centre for Population Neuroscience and Stratified Medicine (PONS), Department of Psychiatry and Neuroscience, Charité Universitätsmedizin Berlin, Berlin, Germany
        $^{w}$School of Psychology and Global Brain Health Institute, Trinity College Dublin, Dublin, Ireland
        $^{x}$Centre for Population Neuroscience and Precision Medicine (PONS), Institute for Science and Technology of Brain-inspired Intelligence (ISTBI), Fudan University, Shanghai, China
        $^{y}$Department of Systems Neuroscience, University Medical Center Hamburg-Eppendorf, Hamburg, Germany
        $^{z}$Department of Neurology and Neurosurgery, McGill University, Montreal, Quebec, Canada
        $^{aa}$Department of Child and Adolescent Psychiatry, Psychotherapy and Psychosomatics, University Medical Center Hamburg-Eppendorf, Hamburg, Germany
        $^{**}$Co-last authors
    }
    \\\\
    Corresponding author: \url{hemforthl@gmail.com}
}

\begin{document}

\maketitle

\begin{abstract}
Incomplete Hippocampal Inversion (IHI), sometimes called hippocampal malrotation, is an atypical anatomical pattern of the hippocampus found in about 20\% of the general population.  IHI can be visually assessed on coronal slices of T1 weighted MR images, using a composite score that combines four anatomical criteria.
IHI has been associated with several brain disorders (epilepsy, schizophrenia). However, these studies were based on small samples. Furthermore, the factors (genetic or environmental) that contribute to the genesis of IHI are largely unknown. Large-scale studies are thus needed to further understand IHI and their potential relationships to neurological and psychiatric disorders.
However, visual evaluation is long and tedious, justifying the need for an automatic method.  
In this paper, we propose, for the first time, to automatically rate IHI. We proceed by predicting four anatomical criteria, which are then summed up to form the IHI score, providing the advantage of an interpretable score. We provided an extensive experimental investigation of different machine learning methods and training strategies. We performed automatic rating using a variety of deep learning models ("conv5-FC3", ResNet and "SECNN") as well as a ridge regression. We studied the generalization of our models using different cohorts and performed multi-cohort learning. We relied on a large population of 2,008 participants from the IMAGEN study, 993 and 403 participants from the QTIM and QTAB studies as well as 985 subjects from the UKBiobank. 
We showed that deep learning models outperformed a ridge regression. We demonstrated that the performances of the "conv5-FC3" network were at least as good as more complex networks while maintaining a low complexity and computation time. We showed that training on a single cohort may lack in variability while training on several cohorts improves generalization (acceptable performances on all tested cohorts including some that are not included in training). The trained models are available at \url{https://github.com/LisaHemforth/AutomaticIHIRating}.
\end{abstract}


\section{Introduction}

Incomplete Hippocampal Inversion is an atypical anatomical pattern found in 15 to 20 percent of the general population with a higher prevalence in the left hemisphere (~20\% compared to ~9\% in the right hemisphere)~\citep{ClaireIHI2015, Bronen91normalhipp, Bernasconi2005healthy, Bajic2008dev}. It can be referred to as "incomplete hippocampal inversion"~\citep{ClaireIHI2015, Bajic2008dev}, "hippocampal malrotation"~\citep{Barsi2000malrot} or "abnormal hippocampal formation"~\citep{Bernasconi2005healthy}. In this work, we will be referring to it as "incomplete hippocampal inversion" (IHI).

Its origins are unclear but IHI are likely to be formed during pre-natal development. It is during this time that most gyri are formed and that the hippocampus is folded~\citep{Bajic2008dev}. Furthermore, the growth of left and right hemispheres react differently to maternal stress~\citep{Qui2013hippdev}. 
IHI is associated with variations of hippocampal subfields, namely smaller CA1 (first region of the cornu Ammonis)~\citep{Colenutt18subfield} and larger subiculum~\citep{Fragueiro23sub}. Furthermore, it has been linked to variations in sulcal patterns~\citep{ClaireIHI2015}. Thus, it may be associated to an overall atypical brain development. 

A previous study has investigated the genetic underpinnings of IHI~\citep{GWAS} and suggested a moderate, statistically significant, heritability ($h^2$ of 0.54). The Genome Wide Association Study (GWAS) did not identify any causal genetic variant due to its limited sample size~\citep{GWAS}. 

Some studies (albeit conducted on small samples) have shown a higher prevalence of IHI, compared to general population, in patients suffering from epilepsy~\citep{Lehericy95epilepsy, Baulac98epilepsy,Bernasconi2005healthy} and schizophrenia~\citep{Roeske21schiz}. Autism spectrum disorder~\citep{ Campbell06delsyn} has also been noted in association with IHI. This is also the case of other structural variations such as agenesis of the corpus callosum~\citep{Atlas86corpcall}.  More generally, we can hypothesise that IHI could be linked to pathologies associated with hippocampal structure. For example, Alzheimer's disease is associated with progressive hippocampal atrophy~\citep{Barnes09Alzheimers}, while patients suffering from schizophrenia showed smaller hippocampi~\citep{VanErp16schizophrenia}. Similar results were found for a variety of psychiatric disorders such as obsessive compulsive disorder (OCD), major depressive disorder (MDD), attention deficit/ hyperactivity disorder (ADHD) or post traumatic stress disorder (PTSD), by the ENIGMA consortium~\citep{Thompson20enigma}. Furthermore, there has been shown to be a link between hippocampal volume in MDD and the response to antidepressants~\citep{Colle16Depression}. Moreover, it has been shown that IHI affects automatic segmentation algorithms: segmentation is less accurate in the presence of IHI~\citep{Kim12seg, Fragueiro23sub}. Thus, it is possible that some studies reporting associations with hippocampal volume are actually confounded by IHI leading to erroneous conclusions that a given disorder, genotype or trait is associated with changes in hippocampal volume while it is in fact associated with IHI.
To date, there are no large-scale studies of IHI allowing to confirm or investigate the links with disorders of the brain, and to progress our understanding of the causes and consequences of IHI. This is, in part, due to the difficulty of visually rating IHI on large cohorts.

Indeed, IHI annotations can prove to be a long and tedious task and visual rating is not adapted to large-scale studies. Using an automatic rating can facilitate this process and make it possible to study large and different cohorts.

In a preliminary conference proceedings paper, we have shown the feasibility of automatically detecting IHI using the previously mentioned annotation protocol using only one linear and one deep learning model~\citep{SPIE_paper}, on one cohort (IMAGEN), the one manually rated in~\citep{ClaireIHI2015}.
In this paper, we extended our previous work by: i) considering a larger scope of machine learning methods including ridge regression, a Conv5-FC3, a ResNet and a squeeze and excite ResNet, which have been shown to be effective in computer vision; ii) evaluating approaches on three additional datasets (QTIM, QTAB, a subsample of the UKBiobank); iii) studying the benefits of multi-cohort training to improve generalization of the automatic rating. 

\section{Data and pre-processing}
\subsection{Manual IHI rating protocol}
\begin{figure} [ht]
  \begin{center}
  \begin{tabular}{c} 
   \includegraphics[width=\textwidth]{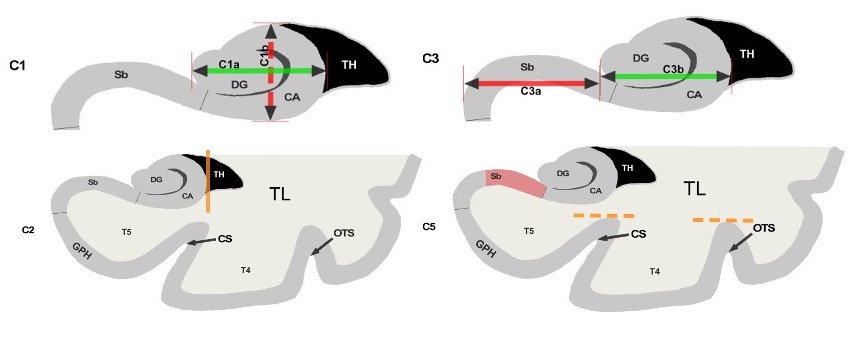}
  \vspace{-0.4cm}
  \end{tabular}
  \end{center}
  \caption[crit] 
  { \label{fig:crit} 
\emph{Schematic of the visual criteria.} 1: Verticality and roundness of the hippocampal body. 2: Verticality and depth of the collateral sulcus. 3: Medial position of the hippocampus. 5: Depth of the collateral sulcus and occipito-temporal sulcus. Reproduced from [1] (CC BY).}
  \end{figure}

We have trained our algorithms against robust manual ratings, using a reproducible annotation protocol presented in~\cite{ClaireIHI2015} which takes into account the most representative criteria of IHI~\citep{Baulac98epilepsy}, keeping a reasonable number of anatomical criteria (five) without overbearing the annotator. These annotations are made on coronal slices of T1 weighted MR images. The first criterion (C1) assesses the verticality and roundness of the hippocampal body. The second criterion (C2) evaluates the verticality and depth of the collateral sulcus. The third criterion (C3) quantifies the medial position of the hippocampus. The fourth criterion (C4) indicates if the subiculum is bulging upwards or not.  The fifth criterion (C5) assesses whether any sulci of the fusiform gyrus exceed the level of the subiculum. The total IHI score is then the sum of the individual criteria. Here, we did not use the criterion C4 because it is very rare (i.e. no bulge in $\ge97\%$ of individuals) and is notoriously difficult to rate  with low test-retest reliability~\citep{ClaireIHI2015}. Note that due to its low frequency, its exclusion has nearly no effect on the total IHI score. Each criterion we considered is rated on a 2-points scale with a step of 0.5 for criteria 1 to 3 and a step of 1 for criterion 5. A visual schematic of these criteria extracted from Cury et al.~\citep{ClaireIHI2015} can be found in Figure~\ref{fig:crit}.

\subsection{Cohorts description}
\paragraph{Subjects: }
We studied 2,008 participants from the multicentric IMAGEN study. We included all participants with a T1-weighted anatomical MRI acquired at 3 Tesla that passed a visual quality check during rating. We used the first acquisition session where participants were 14 years old in average. We also used 993 subjects of the QTIM and 400 subjects of the QTAB cohort, which are both twin studies from Queensland. Finally, a subset of 985 subjects of the UKBiobank was included. All images were checked to be of sufficient quality for visual rating. A short description of these cohorts can be found in table~\ref{tab:cohorts}. The choice of cohorts was performed to increase variability in terms of age and acquisition sequences during training.

\begin{table}[hbtp]
\caption{\emph{Description of cohorts.} Table summarizing the number of images, age in years (shown as mean $\pm$ standard deviation), the rater, the 95\% CI of the prevalence of IHI, the origin of the cohort and the sex distribution.} 
\label{tab:cohorts}
\begin{center} 
\resizebox{\textwidth}{!}{\begin{tabular}{|c|ccccccc|} 
\hline
cohort    & Number of subjects  & Age    & Rater    & \% IHI (left)  & \% IHI (right) & Origin & \% Female  \\
\hline
IMAGEN  & 2008   & 14.5$\pm$1.3   & CC   & [19, 23]    & [6, 10]     &  Europe  & 51.27\\
\hline
QTIM  & 993   & 22.9$\pm$2.8   & KDM   & [21, 27]    & [8, 12] &  Australia & 61.09\\
\hline
QTAB    & 400     & 11.3$\pm$1.3    & KDM      & [22, 32]  & [13, 21]  & Australia & 48.89\\
\hline
UKBiobank    & 985   & 63.5$\pm$7.6  & KDM      & [18, 24]   & [4, 8]  & UK & 58.09\\
\hline
\end{tabular}
}
\end{center}
\end{table}

\paragraph{MRI acquisition: }
IMAGEN was acquired in 8 different acquisition sites in Europe using a variety of 3 Tesla scanners (Siemens Verio and TimTrio, Philips Achieva, General Electric Signa Excite, and Signa HDx). T1-weighted images of the cohort were obtained using an MPRAGE sequence (TR=2300ms; TE=2.8ms; flip angle=9°; resolution=1.1mm×1.1mm×1.1mm).

QTIM was acquired using a 4 Tesla Bruker Medspec scanner using an inversion recovery rapid gradient echo protocol (TI=700ms; TR=1500ms; TE=3.35ms; flip angle=8°; resolution=0.94mm×0.98mm×0.98mm).

QTAB was acquired on a 3 Tesla Magnetom Prisma scanner (Siemens Medical Solutions, Erlangen) using a 3D MP2RAGE  sequence (TI=700ms; TR=4000ms; TE=2.99ms; flip angle=6°; resolution=0.8mm×0.8mm×0.8mm).

The UKBiobank was acquired on 3 Tesla Siemens Skyra scanners in 3 acquisition sites using a 3D MPRAGE sequence (TR=2000ms; TI=880ms; resolution=1mmx1mmx1mm)
 
\subsection{MRI preprocessing}
We processed the MRI using the first setp of the t1-volume pipeline\footnote{\url{https://aramislab.paris.inria.fr/clinica/docs/public/latest/Pipelines/T1_Volume/}} implemented in Clinica~\citep{Routier21, Gonzales18}. This pipeline is a wrapper of the {\sl Segmentation}, {\sl Run Dartel} and {\sl Normalise to MNI Space} routines implemented in SPM. 
During the first step, the Unified Segmentation procedure~\citep{Ashburner05} is used to simultaneously perform tissue segmentation, bias correction and spatial normalization of the input image. Here we use the spatially normalized greymatter maps.

We then cropped images around the hippocampi and close surrounding sulci ([24:96,54:107,16:49] in MNI coordinates). In supplementary material (supplementary figure~\ref{fig:ROI}), we study the impact of the choice of the region of interest (ROIs) and demonstrate that the above choice leads to performances which are at least as good as other choices while being less computationally expensive.

\subsection{IHI annotation on cohorts}
All images were annotated by experts, either Claire Cury (CC) or Kevin de Matos (KDM). To estimate inter and intra-rater variability, 100 images of the IMAGEN cohort were annotated by both raters and twice by rater KDM, several weeks apart. We expect the inter-rater reliability to be the maximal prediction accuracy achievable with an automated method, as it quantifies the amount of uncertainty in the manual rating.  We reported the frequency of IHI in each hemisphere (Table~\ref{tab:cohorts}). Inversion was deemed incomplete when the composite score was greater or equal to 4, in accordance with the threshold recommended previously~\citep{ClaireIHI2015}. 

\section{Methods and analysis}

\subsection{Test/train sets}
\label{sec:sets}
We isolated 25\% of the participants of each cohort to form a test set. We performed the split prior to running any analysis and only used the test set to evaluate results. To ensure that the test set is representative of the full sample we stratified the split based on all IHI criteria as well as age, weight, height, sex, handedness and imaging centre. In practice, we performed 200 random splits and selected the one that minimised differences in distributions for all considered variables between the training and test sets (based on a Kolmogorov-Smirnoff test). 
We tuned hyper-parameters using the remaining data. We further split this data into a training (80\% of individuals) and validation set (20\%). We used the same split rules as above to ensure comparability of all the splits. Table~\ref{tab:testrainsets} shows the amounts of data from each cohort in test, training and validation sets.

\begin{table}[hbtp]
\caption{\emph{Description of sets.} Number of images in train, validation and test sets for each cohort. Note that we varied the number of cohorts included in the training.} 
\label{tab:testrainsets}
\begin{center} 
\begin{tabular}{|c|cccc|} 
\hline
cohort    & Train-set    & Validation-set & Test-set    & Total   \\
\hline
IMAGEN  & 1205   & 301   &  502 & 2008 \\
\hline
QTIM  & 596   & 149   & 248   & 993 \\
\hline
QTAB    & 240     & 60   & 100      & 400  \\
\hline
UKBiobank    & 554   & 185  & 246      & 985 \\
\hline
\end{tabular}
\end{center}
\end{table}

\subsection{Training strategies}
\label{sec:trainMeth}
In order to assess how the models perform in different cohorts and what is the influence of the cohorts used for training, we proceeded with three different training sets and evaluated the predictive ability on the four test sets independently and pooled together.
\begin{itemize}
  \item \emph{IMAGEN training strategy: } First, we trained the models using the IMAGEN training set only. Predictive performance in the test sets from the 3 independent cohorts, QTIM, QTAB and UKB assesses the generalizability of this strategy. A risk of this approach is that IMAGEN is fairly homogeneous (mean age 14.5$\pm$1.3), and training on this unique cohort may lead to over-fitting the sample characteristics or age group, even if IMAGEN used a multi-centric design that used different scanners. 
  \item \emph{IMAGEN, QTIM, QTAB training strategy: } To introduce more variance into the training, we combined the training sets of IMAGEN, QTIM and QTAB. In addition to increasing the training sample size, this introduces new age groups, new scanners and acquisition sequences, as well as a new rater into the training. We kept the UKbiobank as an independent validation cohort to test for generalizability of the prediction.
  \item \emph{ALL training strategy: } Lastly, we performed multi-cohort training including all training sets (IMAGEN, QTIM, QTAB and UKBiobank). This further increases the training sample size and gives us the opportunity to test whether performance on the UKBiobank improves when including a part of the cohort in the training set.

\end{itemize}

\subsection{Deep learning models}
We trained three neural networks which are implemented in ClinicaDL~\citep{thibeau2022clinicadl}\footnote{\url{https://clinicadl.readthedocs.io/en/latest/Train/Introduction/}}, an open source software package for deep learning analysis of neuroimaging data using 3D MRI data cropped around the hippocampus and surrounding sulci:  
\begin{itemize}
    \item \emph{Conv5-FC3 model: } a convolutional neural network made of 5 convolutional blocks and three fully connected layers. Each of the convolutional blocks is made of one convolutional layer, a batch normalization, a ReLu and a Max pooling. This CNN is fairly shallow, easy to train, and has shown good performance at MRI based prediction of Alzheimer's disease~\citep{WEN2020101694}.
    \item \emph{ResNet model: }a 3D ResNet~\citep{Jonsson19resnet} made of five residual blocks separated by a max pooling and a final block composed of a fully connected, a ReLu, a dropout, a concentration layer and a final fully connected layer. We used the default dropout of 0.5. This model has previously been used by our team in~\cite{PAC2019} for brain age prediction, and it is a reference model in computer vision.
    \item \emph{SECNN model: }a squeeze and excite ResNet, which we will be referring to as "SECNN", based on the theory of~\cite{SECNN, Hu18SECNN}. This model is similar to the 3D ResNet, except that it contains an extra squeeze and excite block and ReLu in the residual blocks. Squeeze and excite blocks provide the advantage of improving channel inter-dependencies and have been shown to perform well on computer vision tasks~\citep{Hu18SECNN}.
    
\end{itemize}.

\begin{figure} [ht]
  \begin{center}
  \begin{tabular}{c} 
   \includegraphics[width=\textwidth]{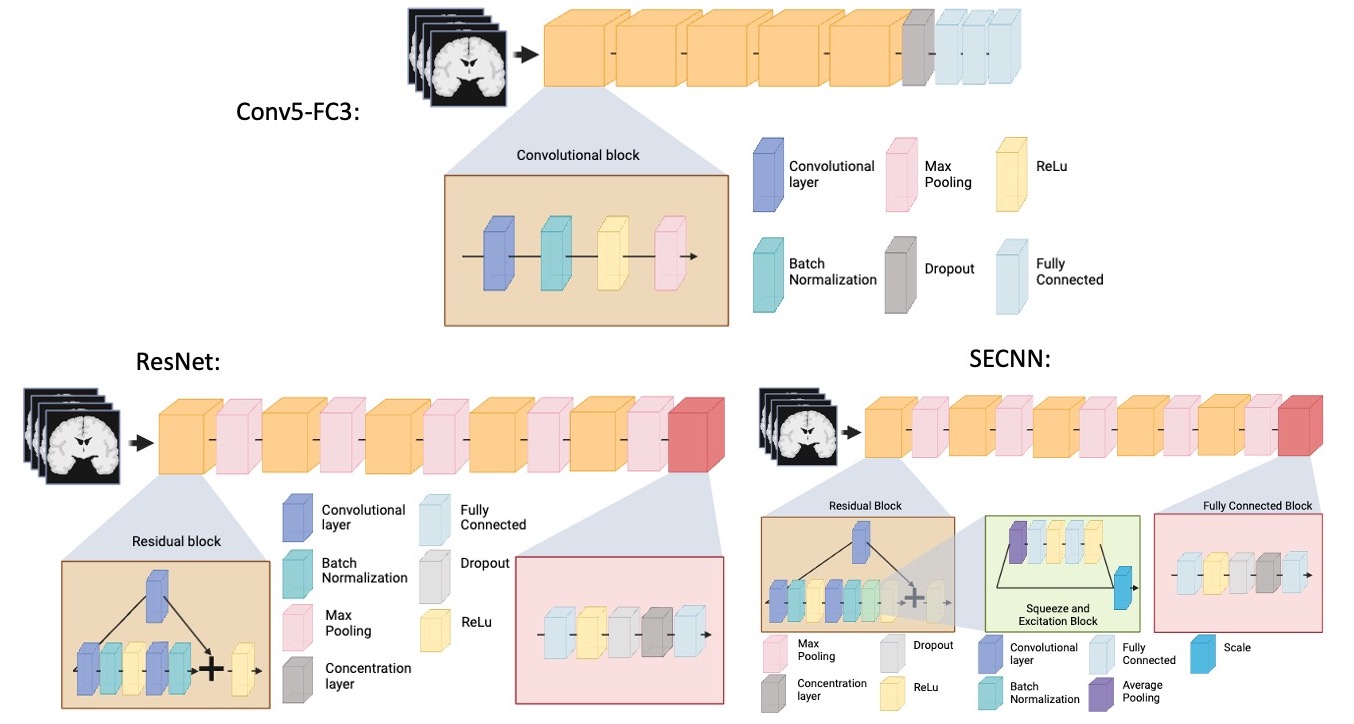}
  \vspace{-0.4cm}
  \end{tabular}
  \end{center}
  \caption[models] 
  { \label{fig:models} 
Schematic representation of the deep learning models used for prediction.}
  \end{figure}

A schematic representation of the previously mentioned models can be found in Figure~\ref{fig:models}. All models were trained for a regression task over a maximum of 50 epochs using the mean squared error as a loss function. The model with the lowest loss on the validation set over the epochs was used for analysis. We used a batch size of 16 to allow for several images per cohort in each batch. We used the Adam optimizer and a learning rate of 1e-4 with a weight decay of 1e-4. The tolerance was set to 0. Models were implemented in Pytorch.

Further attempts at improving the results included data augmentation and oversampling. The implemented methods and results can be found in the supplementary materials.

\subsection{Linear models}
To benchmark the performance of deep-learning models against simpler linear ones, we also performed automatic rating of IHI with a ridge regression. We used the ridge regression implemented in scikit-learn~\cite{pedregosa2011scikit} \footnote{\url{https://scikit-learn.org/stable/modules/generated/sklearn.linear_model.Ridge.html}}. Flattened images were used as input. The hyper-parameter were chosen through nested cross-validation with 5 outer layers and 6 inner layers. The data used over all splits corresponds to the union of training and validation data from Table~\ref{tab:testrainsets}. The splits were performed arbitrarily by the \texttt{KFold} function of scikit-learn.

\subsection{Statistical analysis}
The IHI individual criteria range from 0 to 2 with 0.5 or 1 point steps. Thus, we rounded the predicted scores to the closest 0.5 mark for criteria 1,2 and 3, and the closest unit for criterion 5 to correspond to the human ratings. We constructed the (predicted) global IHI scores by summing the prediction of each IHI criterion. In the following, they are denoted as 'SCi{\_L or \_R}\_add' for the left and right hemispheres.

We used Intraclass Correlations (ICC) to evaluate prediction of the global IHI score, which are nearly continuous. For this, we used the \texttt{intraclass\_corr} function implemented in the {\sl pingouin} package ~\cite{vallat2018pingouin}. We used Cohen's Kappa score to evaluate the prediction of individual criteria. For criteria 1-3, which are ordinal, we used a quadratically weighted Kappa. For criterion 5 (0-1 score), we used a standard Kappa. We used the \texttt{cohen\_kappa\_score} implemented in {\sl sklearn.metrics}.

We derived 95\% confidence intervals and standard-error (SE) of the prediction accuracy using a bootstrap approach. The bootstrap was performed using 100 iterations, each consisting of drawing N samples with replacement from the test-set, N being the size of the test set. 

To statistically assess the difference in performance between methods, we then computed the difference of metrics obtained on the same bootstrap samples using different methods. In other words, we obtain a bootstrap of the difference in performance between two given strategies. The mean and standard error obtained from this bootstrap are then used to perform a Student's t-test. We use a Bonferroni correction on p-values obtained on each criterion or composite score. We define statistical significance as corrected p-value<0.05.

\section{Results}
\subsection{Model and training set performances}

\begin{figure} [ht]
  \begin{center}
  \begin{tabular}{c} 
   \includegraphics[width=\textwidth]{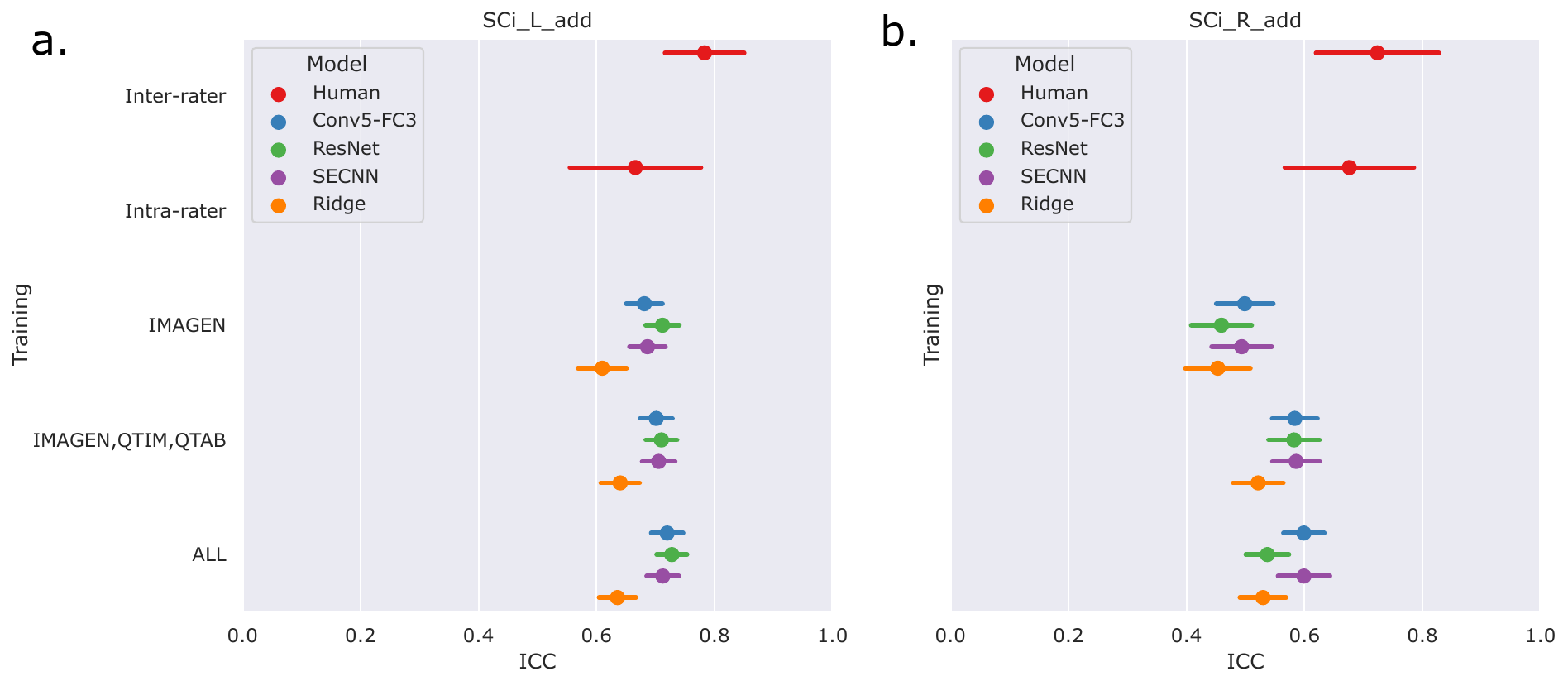}
  \vspace{-0.4cm}
  \end{tabular}
  \end{center}
  \caption[indQTIM] 
  { \label{fig:CompPooled} 
   \emph{Results of the predictions of composite scores on pooled independent test sets of the IMAGEN, QTIM, QTAB and UKB cohorts.}  We show the mean ICC and 95\% confidence intervals obtained through bootstrapping. Results are shown for the three assessed deep learning models (Conv5-FC3, ResNet and SECNN) and the ridge regression, alongside inter-rater and intra-rater performances. Three training strategies are compared (IMAGEN strategy, IMAGEN, QTIM, QTAB strategy and ALL strategy). These results are shown for predictions in the left (panel a) and right (panel b) hemispheres.} 
  \end{figure}

We first examine the performances of composite score predictions of each model on a pooled test set of all cohorts (N=502+248+100+245), comparing the results of the three different training strategies (IMAGEN strategy, IMAGEN, QTIM, QTAB strategy and ALL strategy). Results are displayed in Figure~\ref{fig:CompPooled}. Human performances (inter and intra-rater ICCs) are plotted for reference. 

Human performances exhibit large confidence intervals (95\%CI inter-rater ICC = [0.701, 0.849], 95\%CI intra-rater ICC = [0.553, 0.772]) due to the low sample size for their computation (100 images). Inter-rater performances can  be considered as the maximal prediction achievable. It can be noted that deep learning models in the left hemisphere were not deemed different from inter and intra-rater performances (corrected $ p > 0.05$ in all cases). This may be mainly attributed to the large confidence intervals for human performances. However, performances on the left hemisphere (95\% CI ICC = [0.678, 0.729] for 'Conv5-FC', IMAGEN,QTIM,QTAB strategy) remain closer to inter-rater performances than in the right hemisphere (95\% CI ICC = [0.546, 0.620] for 'Conv5-FC3', IMAGEN,QTIM,QTAB strategy). Predictions can hence still be improved in the right hemisphere, in which the lower number of IHI makes it a difficult task to learn. We tested oversampling and data augmentation by flipping images to obtain as many IHI on the right and on the left side, however this did not improve results (see Supplementary material)

The ridge regression showed significantly worse performances in the left hemisphere compared to the deep learning models (corrected $ p < 0.05$ for all tests). We observed this result for all training sets (Figure~\ref{fig:CompPooled}a). In the right hemisphere, the performance of the ridge regression seemed slightly lower than that of deep learning algorithms. This difference was statistically significant in the case of the IMAGEN, QTIM, QTAB strategy for all models and for the CNN and SECNN using the ALL strategy. Due to their greater performance, we are focusing on deep learning models in our subsequent analyses. Particularly, as the Conv5-FC3 is not significantly outperformed by more complex models, which require more computation power, we will focus on the results obtained with this model.

In the left hemisphere, increasing the training sample did not appear to significantly improve IHI prediction. On the contrary, in the right hemisphere, performances improved significantly when extending the training set to QTIM and QTAB for the Conv5-FC3 (corrected $ p < 0.05$). Adding some UKBiobank images into the training did not significantly improve the prediction performance, in either hemisphere. For completeness, we have reported the results (Figure~\ref{fig:CompositeALL}) for each specific test-set (IMAGEN, QTIM, QTAB or UKBiobank). A similar pattern of results emerge.

\subsection{Performance on individual criteria}
\begin{figure} [ht]
  \begin{center}
  \begin{tabular}{c} 
   \includegraphics[width=\textwidth]{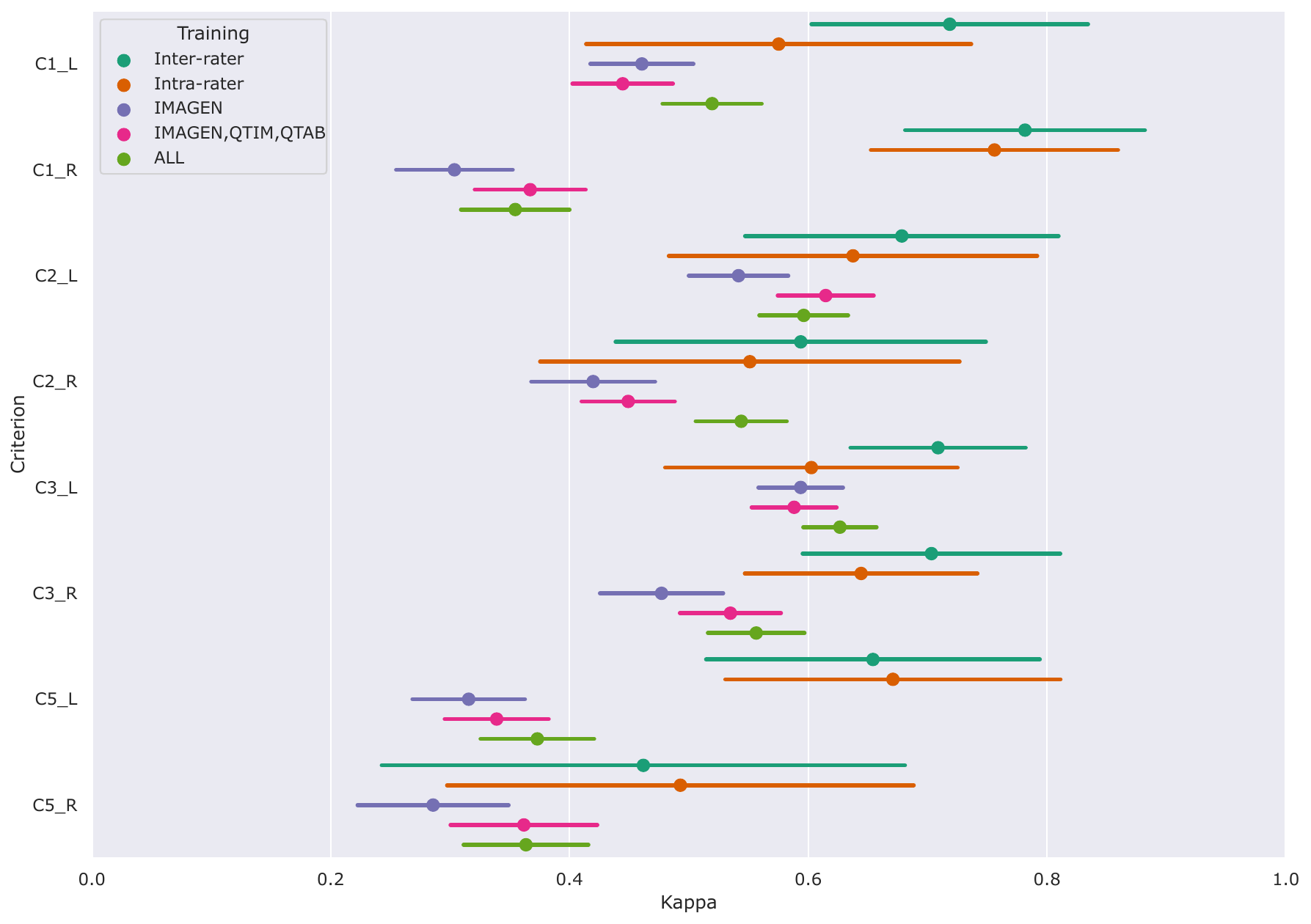}
  \vspace{-0.4cm}
  \end{tabular}
  \end{center}
  \caption[indQTIM] 
  { \label{fig:IndivPooled} 
   \emph{Results of the predictions of individual criteria on pooled independent test sets of the IMAGEN, QTIM, QTAB and UKB cohorts.} We show the mean metrics (weighted kappas for C1 C2 and C3, and an unweighted kappa for C5) and 95\% confidence intervals obtained through bootstraping. Results are shown for the Conv5-FC3, alongside inter-rater and intra-rater performances. Three training methods are compared: using only the training set of the IMAGEN cohort, using the training sets of IMAGEN, QTIM and QTAB  cohorts and using the training sets of all  cohorts (IMAGEN, QTIM, QTAB, UKBiobank).} 
  \end{figure}

Figure~\ref{fig:IndivPooled} shows the performances of the Conv5-FC3 for the prediction of individual criteria.

In the left hemisphere, the performance overlaps with  inter or intra-rater reliability for most criteria (corrected $p > 0.05$). However, this is not the case for C5 (corrected $p < 0.05$ in all cases). This criterion is particular as it is not linear but is still estimated by a regression. We also attempted at using a classifier (see results for RidgeClass and RidgeClassOS on~\ref{fig:linearmodels}) but this did not prove fruitful.

Compared to the left, performances in the right hemisphere are in general lower.  
As for the composite score, this is likely due to the lower prevalence of IHI in individual scores, as it was already shown in~\cite{ClaireIHI2015}. 

\subsection{Visual analysis of the trained networks}
\begin{figure} [ht]
  \begin{center}
  \begin{tabular}{c} 
   \includegraphics[width=\textwidth]{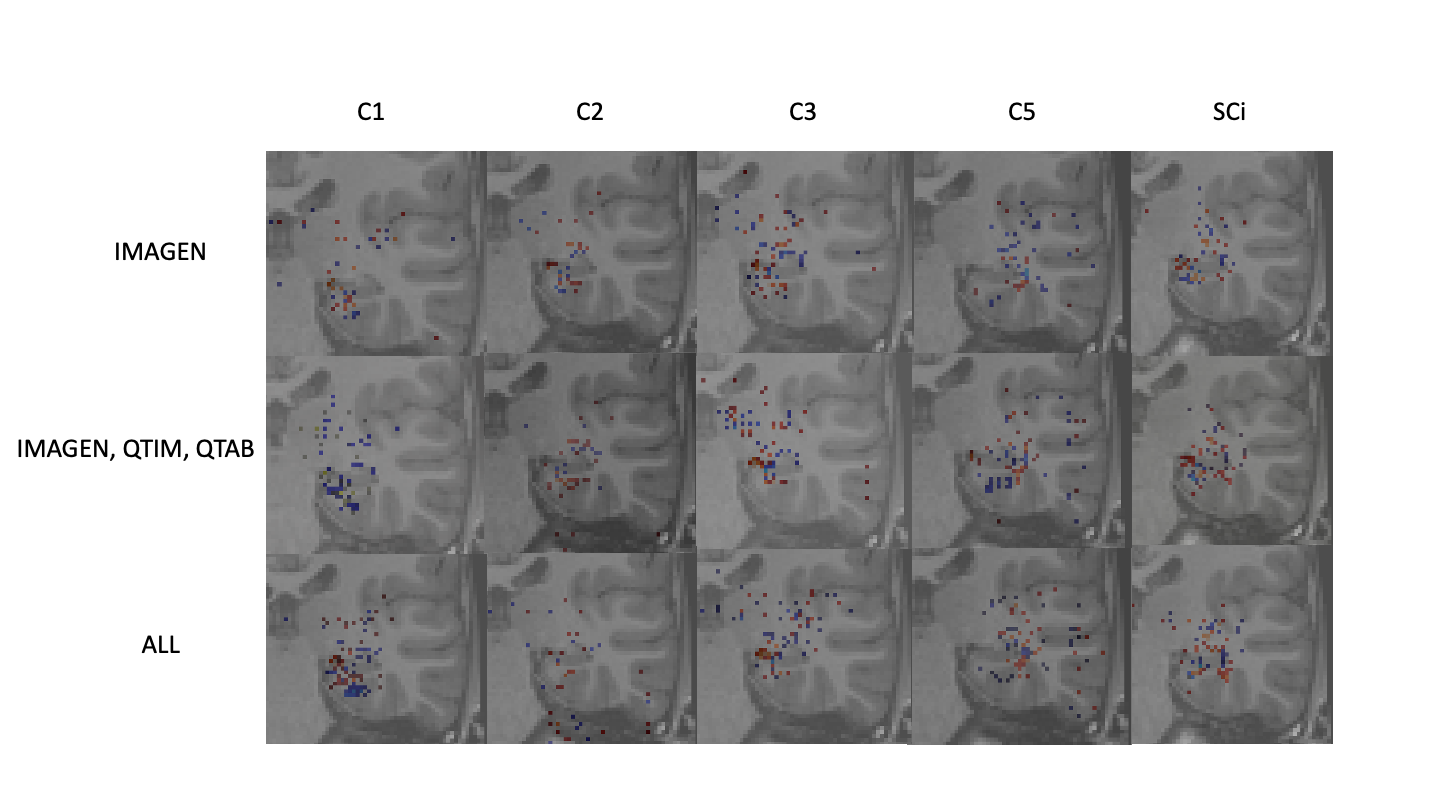}
  \vspace{-0.4cm}
  \end{tabular}
  \end{center}
  \caption[crit] 
  { \label{fig:saliency} 
  \emph{Saliency maps} extracted from the Conv5-FC3 model's predictions on the UKBiobank in the left hemisphere. Plots are shown for all training strategies, for individual criteria and the composite scores. Saliency maps were thresholded to show only the highest weights and overlayed on a T1 weighted MRI image.} 
  \end{figure}

As a sanity check, some group saliency maps~\citep{Simonyan13saliency} were extracted from Conv5-FC3 nets as implemented in ClinicaDL~\citep{thibeau2022clinicadl}. The maps were obtained through back-propagation and are shown for all three training sets with various training strategies of data in figure~\ref{fig:saliency}.

While saliency maps are limited in their analysis~\citep{Saliency}, they can serve as a sanity check of our processes. Here we show only the 1000 highest values to ensure visual coherence. Figure~\ref{fig:saliency} shows that weights are mostly concentrated on the hippocampus and surrounding regions for all training methods and criteria. Some criteria show sparser maps, such as C2 predictions trained on all data, but no maps show a complete absence of weights in the region of interest. Composite scores predicted with a model trained on several cohorts show maps with weights centered around the hippocampus. These results show that our networks are indeed using hippocampus-related features.

\section{Discussion}
Our main goal for this study is to establish an automatic rating method for IHI. To ensure generalization, we performed multi-cohort training using different strategies. The results emerging from this study are that deep learning models outperform the linear method. This makes sense as we are assessing complex aspects of hippocampus structure and shape which may be hard to predict from a linear combination of voxel intensities. Another observation was that all deep learning models performed similarly, although based on the size of our test sample, we could only significantly detect differences of 0.125 ICC points (at 80\% power).  This suggests that a simpler model may suffice for this task, which has the advantage of requiring less computation, and is easier to share and utilise. We would recommend using our "Conv5-FC3" trained network for IHI prediction. IHI prediction in the left hemisphere approached that achieved from human raters. On the other hand, automatic rating of IHI in the right hemisphere, remained below the automatic rating performances in the left hemisphere, suggesting our algorithms could be improved. 

We then looked into the performances of the models when trained with only the IMAGEN cohort, IMAGEN, QTIM and QTAB and finally with all available cohorts. Using only IMAGEN proved to be significantly less performing than using several cohorts when analysing results over all cohorts in the right hemisphere. This may be due to a lack of variability to accommodate new cohorts when using only IMAGEN for training. By adding different cohorts into the training-set we increase its variability in terms of raters, age and acquisition sequence. Indeed, results were improved when adding QTIM and QTAB into the training. However, we did not observe further improvement when adding UKBiobank participants. Of course, we cannot rule out that training with a larger sample of UKBiobank images would have led to higher accuracy. 

To investigate the origin as well as the anatomical interpretability of our results, we have considered individual criteria predictions. All individual criteria, except for C5 have confidence intervals overlapping with those of human performances in the left hemisphere. As mentioned previously, this may however be due to the large confidence intervals of human performances. Unfortunately, the drop in performance could also be noticed for individual criteria prediction for the right hemisphere. Some criteria may prove more difficult to predict and low frequencies suggest that more data may be needed to reach similar performance levels.  C5 is also harder to predict due to it's non-ordinal nature. 

In summary, for automatic rating, we recommend using the "Conv5-FC3" network, as no significant improvements were found with more complex models. This model hence ensures performance while maintaining a low computational cost. As per the ideal training set, it may depend on the use case. If the objective is to predict the global IHI score and use it for further analysis on its own, the model trained on IMAGEN, QTIM and QTAB may provide sufficient performances. However, to maximise the prediction of specific IHI criteria, it may be beneficial to retrain the model with a subset of the new cohort to be rated. However, the improvement may be limited to a handful of criteria for the right hemisphere. In case of future studies about IHI prediction being performed on the UKBiobank, we recommend using the training approach using IMAGEN, QTIM and QTAB as training sets to facilitate testing without having to sort out the data used for training. 
While our study is, to our knowledge, the first to show robust, better than chance IHI prediction, it can still be improved, especially in the right hemisphere.

We obtain lower prediction accuracy in the right hemisphere than in the left for both composite IHI scores and individual criteria. We attributed this to the lower frequency of IHI, although it could also be a more complex task. Indeed, IHI are much rarer in the right hemisphere compared to left. The fact that IHI are rare in the right hemisphere has been widely documented in the literature~\citep{Bajic2008dev,ClaireIHI2015, Colenutt18subfield,Roeske21schiz,fu_hippocampal_2021}. To counter this, data-augmentation and over-sampling methods are often used. While we attempted such techniques, they did not prove efficient in our study. In the case of unilateral right sided IHI, our method remains limited. However, as unilateral right sided IHI are rare, we deem that detecting IHI in the left hemisphere automatically could lead to a facilitated manual annotation in the right hemisphere. 

In the future, we could imagine using our method as a semi-automatic annotation tool  by using the predictions as a first estimate that would then be refined by an expert user. Indeed, as our automatic annotation method provides detailed scores for individual criteria, the method remains interpretable. An expert can hence use provided predictions to estimate which are the subjects of interest and perform fast manual correction by looking into the prediction of each of the criteria. Note nevertheless, that this is not the main objective of the present work which purpose is to be able to annotate large datasets in a fully automatic manner to perform population studies such as genome-wide association studies (GWAS) for instance. In this context, it is acceptable if there remains some noise in the automatic annotation, this will only lead to diminished statistical power for population studies, such power being compensated by the ability to annotate very large datasets. 

To improve performances, it may be interesting to look into data augmentation, particularly in the right hemisphere. As performances are better on non IHI subjects ($SCi < 4$), than on IHI subjects (corrected $p < 0.05$), it may be possible to retrain the model using only IHI images from the new data-set to optimise performances. As retraining the models with part of one's own data-set may prove to be difficult since an access to IMAGEN, QTIM and QTAB is needed, retraining only the top layers of our models could be a solution. Furthermore, self supervised methods such as contrastive learning have been shown to work on neuro-imaging tasks~\citep{dufumier_contrastive_2021}. However, such approaches can be complex as they require vast amounts of data (above 10.000 images), so attention needs to be paid as to which cohorts are used in future studies. This could also be investigated in a future study. 

We chose to rely on a deep learning approach taking voxel-based inputs. Another option could have been to first perform hippocampal segmentation and then use characteristics from this segmentation (either explicitly defined features or  latent features obtained when training an hippocampal segmentation network) for IHI prediction. However, IHI have been shown to affect the accuracy of hippocampal segmentation (segmentation is less accurate in participants with IHI)~\citep{Kim12seg, Fragueiro23sub}. This is the reason why we did not pursue this avenue. However, it could be interesting to compare such an approach to our method in future work.

Our study has the following limitations.
The sparsity of available annotated data led to our models being trained using only the four cohorts mentioned above. Using larger, more diverse datasets could significantly improve the performances. Furthermore, our models were trained using only gradient echo acquired T1 images and Caucasian subjects. This lack in variability means that we cannot ensure performances on images with different acquisition parameters for example. Results should be treated accordingly
First, some noise might be present in the ground truth labels. While consensus rating is ideal, it was not realistic considering the rating time. We, to our knowledge, have the largest annotated sample of data for IHI.
However, human performances (inter and intra-rater reliability) were estimated only on a 100 image sample of IMAGEN.
It is possible that these vary across cohorts. This could provide an explanation as to why performances plateau in the left hemisphere even when adding additional data.
Our data was annotated by two users. More robustness to variability could be introduced by adding more raters to the training. However, more data is not available at this time. We encourage future users who might have access to more annotations to perform a further training step which may include fine-tuning of our pre-trained models.
Furthermore, right hemisphere performances may still be improved. This could simply be due to the low frequency of IHI in this hemisphere.
Generalizability to unseen sequences, machines or specific populations (e.g. disease groups) has not been investigated and is hence not guaranteed.
The power is overall limited to detect small differences in performance. Larger test samples may give a clearer idea of the relative performances of models and training samples. 


\section{Declarations}

\subsection{Availability of data and materials}

Code used to process the data and perform the analyses is available at https://github.com/LisaHemforth/AutomaticIHIRating. 

The QTIM and QTAB dataset are in open access and available online at https://openneuro.org/datasets/ds004169 and https://openneuro.org/datasets/ds004146. 

The IMAGEN dataset is available to interested researchers upon application to the IMAGEN Executive Committee (ponscentre@charite.de, https://imagen-project.org/?page\_id=547).

This research has been conducted using the UK Biobank Resource under Application Number 53185.

\subsection{Author Contributions}

Lisa Hemforth had full access to all the data in the study and takes responsibility for the integrity of the data and the accuracy of the data analysis.    

Study concepts and study design: Lisa Hemforth, Baptiste Couvy-Duchesne, Claire Cury, Olivier Colliot.

Acquisition, analysis or interpretation of data interpretation: all authors.

Manuscript drafting or manuscript revision for important intellectual content: all authors.

Approval of final version of submitted manuscript: all authors.

Literature research: Lisa Hemforth.

Statistical analysis: Lisa Hemforth.

Obtained funding: Baptiste Couvy-Duchesne, Claire Cury, Olivier Colliot.

Administrative, technical, or material support:  Kevin de Matos, Camille Brianceau, Matthieu Joulot, Alexandre Martin.

Study supervision: Lisa Hemforth, Baptiste Couvy-Duchesne, Claire Cury, Olivier. Colliot

 All other authors are part of the IMAGEN consortium. 

\subsection{Declaration of Competing Interests and Funding}

\subsubsection{Disclosure statement}

Competing financial interests related to the present article: none to disclose for all authors.

Competing financial interests unrelated to the present article: OC reports having received consulting fees from AskBio (2020) and Therapanacea (2022-2024), and that his laboratory has received grants (paid to the institution) from Qynapse (2017-2022). Members from his laboratory have co-supervised a PhD thesis with Qynapse (2017-2022). OC’s spouse was an employee of myBrainTechnologies and is an employee of DiamPark. OC holds a patent registered at the International Bureau of the World Intellectual Property Organization (PCT/IB2016/0526993, Schiratti J-B, Allassonniere S, Colliot O, Durrleman S, A method for determining the temporal progression of a biological phenomenon and associated methods and devices) (2017).

Tobias Banaschewski served in an advisory or consultancy role for eye level, Infectopharm, Lundbeck, Medice, Neurim Pharmaceuticals, Oberberg GmbH, Roche, and Takeda. He received conference support or speaker’s fee by Janssen, Medice and Takeda. He received royalities from Hogrefe, Kohlhammer, CIP Medien, Oxford University Press; the present work is unrelated to these relationships.

Dr. Barker has received honoraria from General Electric Healthcare for teaching on scanner programming courses. Dr. Poustka served in an advisory or consultancy role for Roche and Viforpharm and received speaker’s fee by Shire. She received royalties from Hogrefe, Kohlhammer and Schattauer. The present work is unrelated to the above grants and relationships.The other authors report no biomedical financial interests or potential conflicts of interest.

\subsubsection{Funding}

The research leading to these results has received funding from the French government under management of Agence Nationale de la Recherche as part of the "Investissements d'avenir" program, reference ANR-19-P3IA-0001 (PRAIRIE 3IA Institute) and reference ANR-10-IAIHU-06 (Agence Nationale de la Recherche-10-IA Institut Hospitalo-Universitaire-6). BCD is supported by INRIA and a CJ Martin fellowship (NHMRC app 1161356).
The Imagen study is supported by the following sources: the European Union-funded FP6 Integrated Project IMAGEN (Reinforcement-related behaviour in normal brain function and psychopathology) (LSHM-CT- 2007-037286), the Horizon 2020 funded ERC Advanced Grant ‘STRATIFY’ (Brain network based stratification of reinforcement-related disorders) (695313), Human Brain Project (HBP SGA 2, 785907, and HBP SGA 3, 945539), the Medical Research Council Grant 'c-VEDA’ (Consortium on Vulnerability to Externalizing Disorders and Addictions) (MR/N000390/1), the National Institute of Health (NIH) (R01DA049238, A decentralized macro and micro gene-by-environment interaction analysis of substance use behavior and its brain biomarkers), the National Institute for Health Research (NIHR) Biomedical Research Centre at South London and Maudsley NHS Foundation Trust and King’s College London, the Bundesministeriumfür Bildung und Forschung (BMBF grants 01GS08152; 01EV0711; Forschungsnetz AERIAL 01EE1406A, 01EE1406B; Forschungsnetz IMAC-Mind 01GL1745B), the Deutsche Forschungsgemeinschaft (DFG grants SM 80/7-2, SFB 940, TRR 265, NE 1383/14-1), the Medical Research Foundation and Medical Research Council (grants MR/R00465X/1 and MR/S020306/1), the National Institutes of Health (NIH) funded ENIGMA (grants 5U54EB020403-05 and 1R56AG058854-01), NSFC grant 82150710554 and European Union funded project ‘environMENTAL’, grant no: 101057429. Further support was provided by grants from: - the ANR (ANR-12-SAMA-0004, AAPG2019 - GeBra), the Eranet Neuron (AF12-NEUR0008-01 - WM2NA; and ANR-18-NEUR00002-01 - ADORe), the Fondation de France (00081242), the Fondation pour la Recherche Médicale (DPA20140629802), the Mission Interministérielle de Lutte-contre-les-Drogues-et-les- Conduites-Addictives (MILDECA), the Assistance-Publique-Hôpitaux-de-Paris and INSERM (interface grant), Paris Sud University IDEX 2012, the Fondation de l’Avenir (grant AP-RM-17-013 ), the Fédération pour la Recherche sur le Cerveau; the National Institutes of Health, Science Foundation Ireland (16/ERCD/3797), U.S.A. (Axon, Testosterone and Mental Health during Adolescence; RO1 MH085772-01A1) and by NIH Consortium grant U54 EB020403, supported by a cross-NIH alliance that funds Big Data to Knowledge Centres of Excellence.
The QTIM study was supported by the National Institute of Child Health and Human Development (R01 HD050735), and the National Health and Medical Research Council (NHMRC 486682, 1009064), Australia.
The QTAB study was funded by the National Health and Medical Research Council (NHMRC APP1078756), Australia. The QTAB study acknowledges the Queensland Twin Registry Study (https://www.qimrberghofer.edu.au/study/queensland-twin-registry-study) for generously sharing database information for recruitment. The QTAB study was further facilitated through access to Twins Research Australia, a national resource supported by a Centre of Research Excellence Grant (ID: 1078102) from the National Health and Medical Research Council.
We acknowledge access to the facilities and expertise of the CIBM Center for Biomedical Imaging, a Swiss research center of excellence founded and supported by Lausanne University Hospital (CHUV), University of Lausanne (UNIL), Ecole polytechnique fédérale de Lausanne (EPFL), University of Geneva (UNIGE) and Geneva University Hospitals (HUG).

\subsection{Ethical Approval}
IMAGEN was approved by the local ethic committees and a detailed description of recruitment, assessment procedures, and exclusion/inclusion criteria have been published in (the IMAGEN consortium et al. 2010), QTIM was approved by the Human Research Ethics Committees (HREC) at the University of Queensland (de Zubicaray et al. 2008), and QTAB was approved by the Children’s Health Queensland HREC and the University of Queensland HREC (Strike et al. 2022b). Informed consent was obtained from all UK Biobank participants. Procedures are controlled by a dedicated Ethics and and Guidance Council (http://www.ukbiobank.ac.uk/thics), with the Ethics and Governance Framework available at https://www.ukbiobank.ac.uk/media/0xsbmfmw/egf.pdf. IRB approval was also obtained from the North West Multi-centre Research Ethics Committee. 

\bibliography{MELBA-Sub/report}


\clearpage

\renewcommand{\theHsection}{A\Alph{section}}

\renewcommand\theHfigure{A{\arabic{figure}}} 
\renewcommand\thefigure{A{\arabic{figure}}} 
\setcounter{figure}{0} 
\renewcommand\theHtable{A\arabic{table}} 
\renewcommand\thetable{A\arabic{table}} 
\setcounter{table}{0}   

\appendix
\section{Automatic rating  methods of incomplete hippocampal inversions on various cohorts: supplementary material} 
\subsection{Supplementary Figures}

\begin{figure}[!htb]
  \begin{center}
  \begin{tabular}{c} 
   \includegraphics[width=\textwidth]{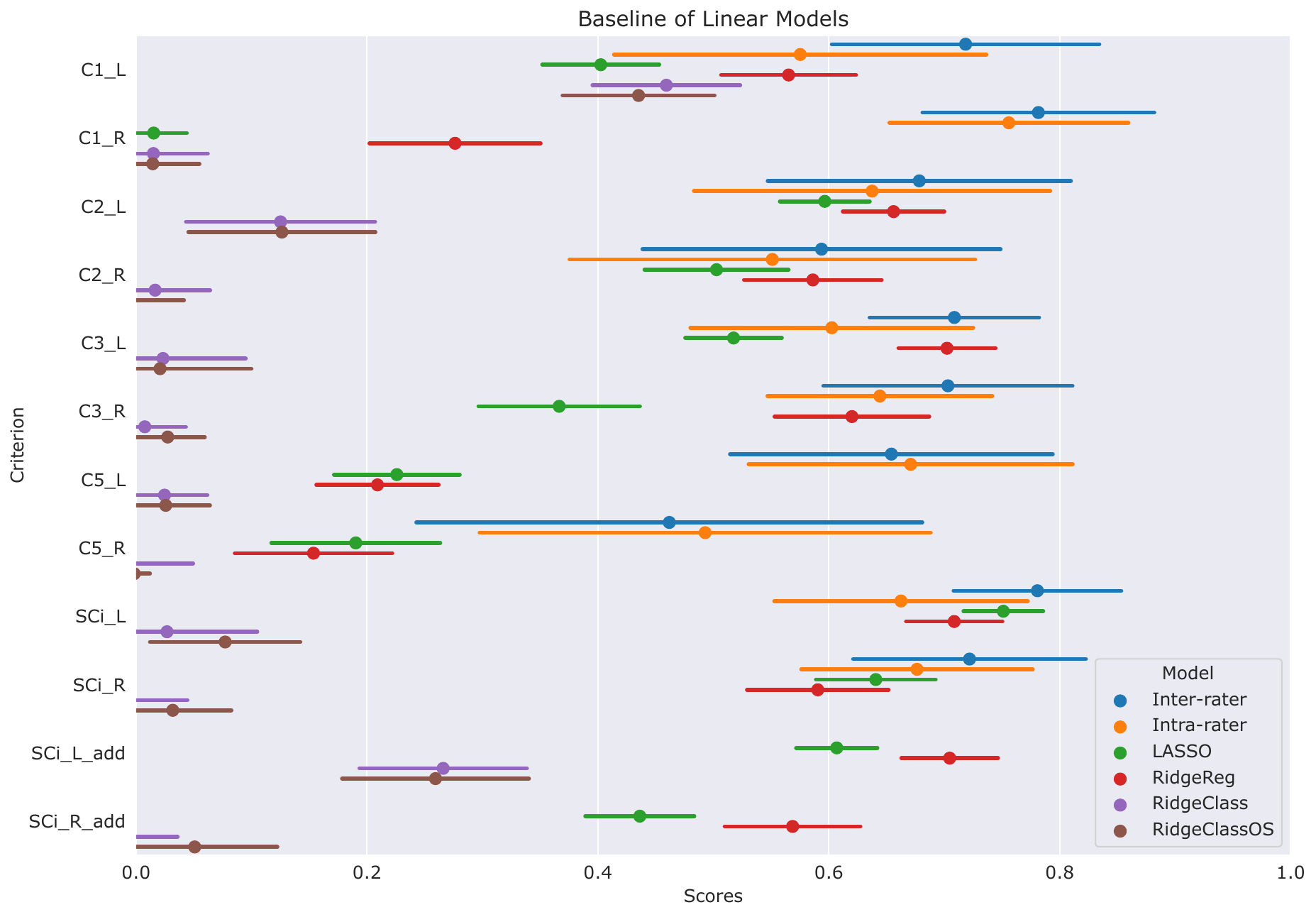}
  \vspace{-0.4cm}
  \end{tabular}
  \end{center}
  \caption[linear] 
  { \label{fig:linearmodels} 
\emph{Results obtained with additional linear models.} LASSO regression, ridge logistic regression classifier (denoted as RidgeClass), ridge logistic regression classifier with oversampling of the minority class (denoted as RidgeClassOS) were studied in addition to the ridge regression (denoted as Ridge) which is presented in the main manuscript. The figure displays the results of the predictions of individual criteria and composite scores on an independent test set of the IMAGEN database.  We show the mean metrics (weighted Cohen's Kappa score for C1, C2 and C3, unweighted Cohen's Kappa score for C5 and ICC for composite scores) and 95\% confidence intervals obtained through bootstraping.}
  \end{figure}
\pagebreak
\begin{figure} [!htb]
  \begin{center}
  \begin{tabular}{c} 
   \includegraphics[width=\textwidth]{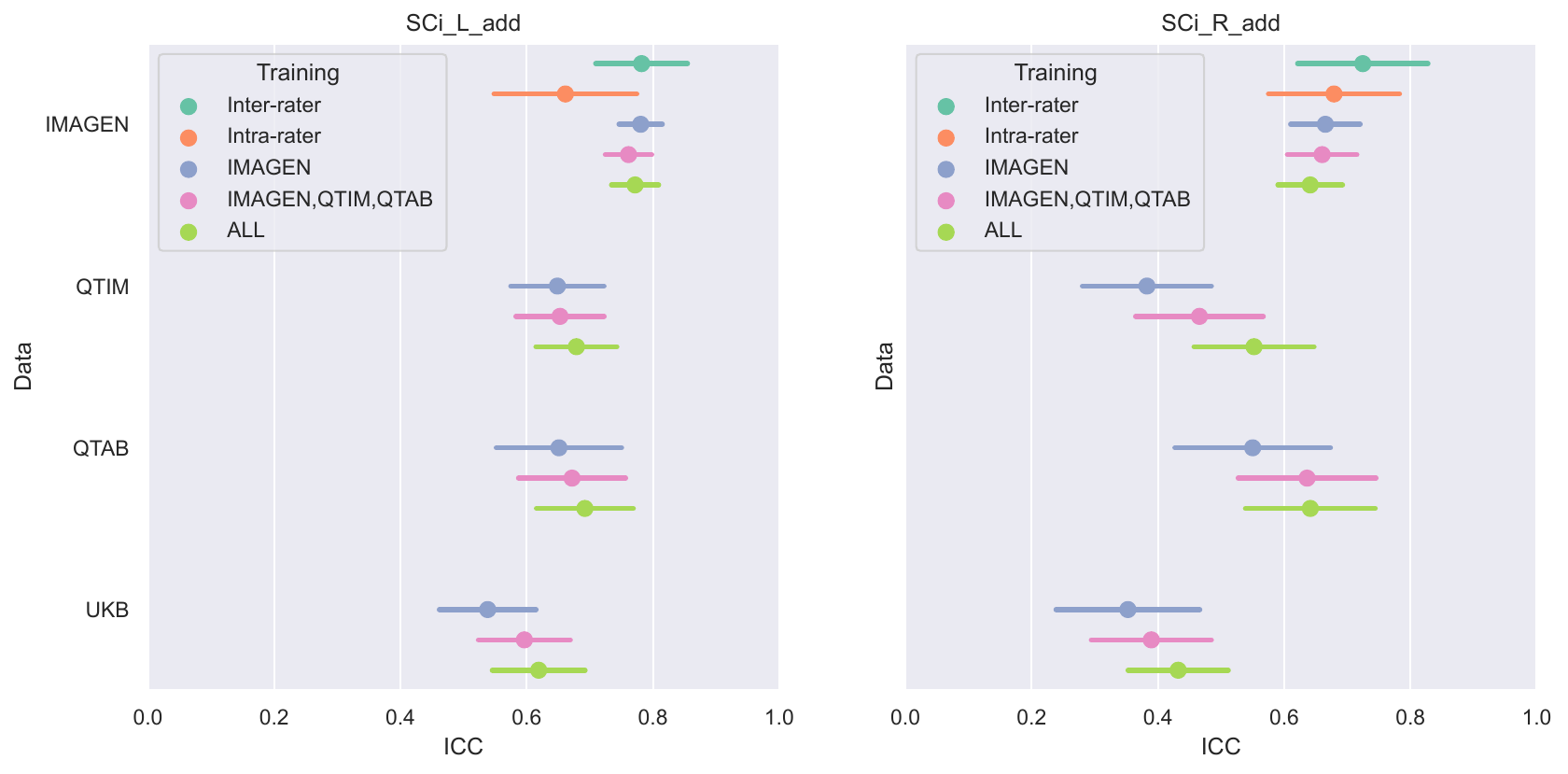}
  \vspace{-0.4cm}
  \end{tabular}
  \end{center}
  \caption[CompAll] 
  { \label{fig:CompositeALL} 
\emph{Results of the predictions of composite scores shown separately on independent test sets of IMAGEN, QTIM, QTAB and UKBiobank.}  We show the mean ICC and 95\% confidence intervals obtained through bootstraping. Results are shown for the Conv5-FC3. Three training methods are compared: using only the training set of the IMAGEN database, using the training sets of IMAGEN, QTIM and QTAB  databases and using the training sets of all databases (IMAGEN, QTIM, QTAB, UKBiobank).}
  \end{figure}
  
\pagebreak
\begin{figure}[!htb]
  \begin{center}
  \begin{tabular}{c} 
   \includegraphics[width=\textwidth]{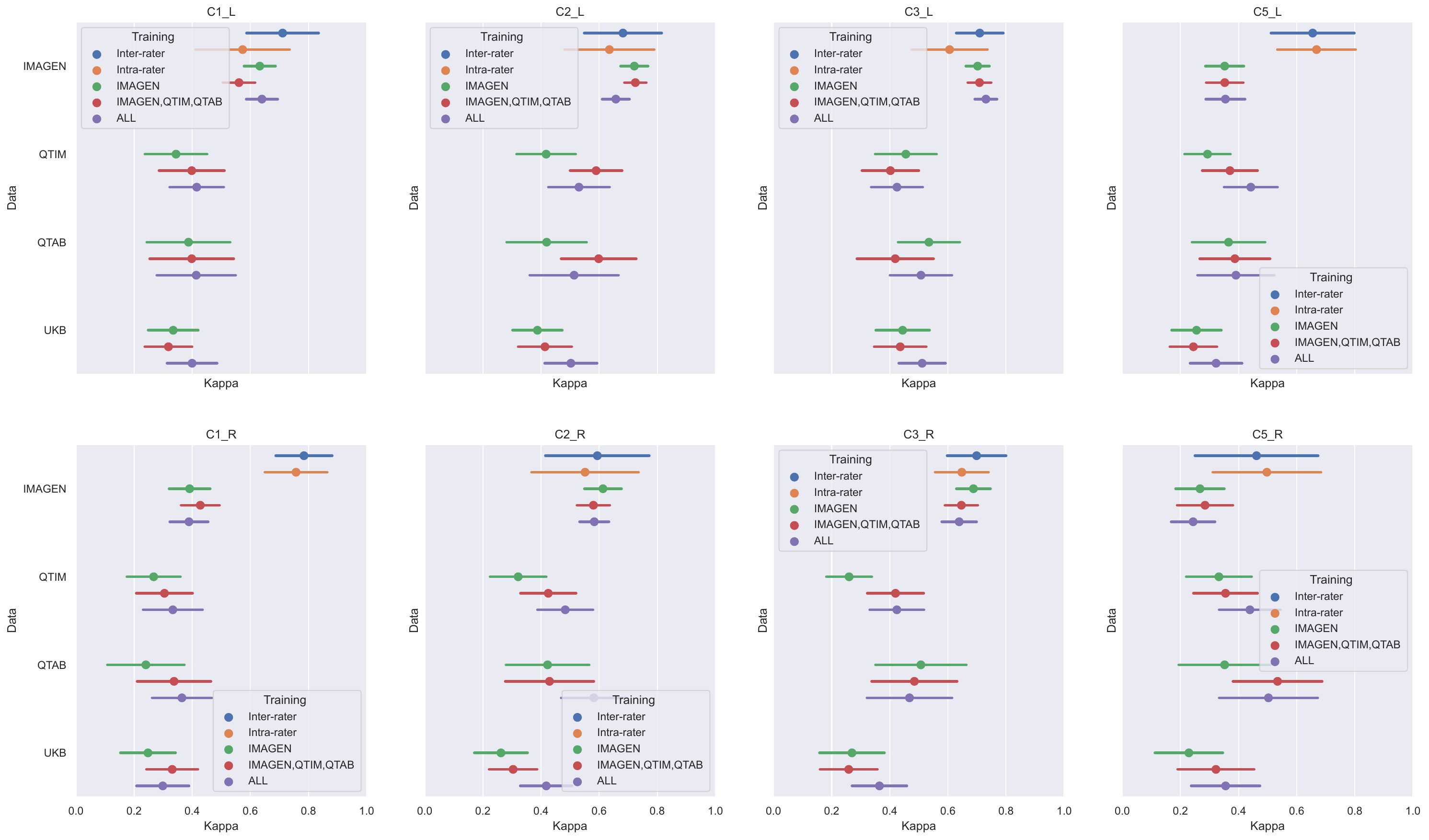}
  \vspace{-0.4cm}
  \end{tabular}
  \end{center}
  \caption[QTAB] 
  { \label{fig:IndivALL} 
\emph{Results of the predictions of individual criteria shown separately on independent test sets of IMAGEN, QTIM, QTAB and UKBiobank.}  We show the mean metrics (weighted Cohen's Kappa score for C1, C2 and C3, unweighted Cohen's Kappa score for C5) and 95\% confidence intervals obtained through bootstraping. Results are shown for the Conv5-FC3. Three training methods are compared: using only the training set of the IMAGEN database, using the training sets of IMAGEN, QTIM and QTAB  databases and using the training sets of all databases (IMAGEN, QTIM, QTAB, UKBiobank).}
  \end{figure}

\begin{figure} [htb]
  \begin{center}
  \begin{tabular}{c} 
   \includegraphics[width=\textwidth]{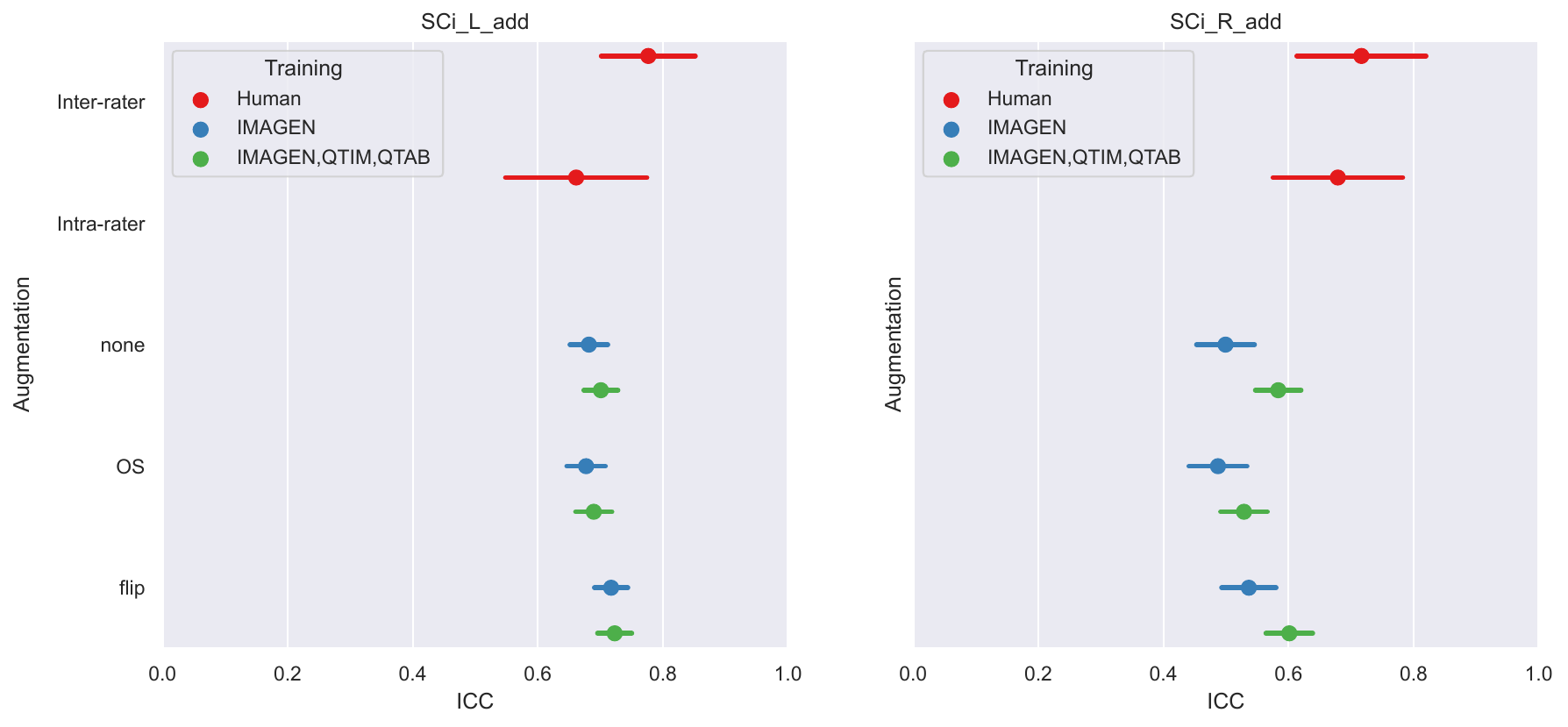}
  \vspace{-0.4cm}
  \end{tabular}
  \end{center}
  \caption[CompPoolAug] 
  { \label{fig:CompositePoolAug} 
\emph{Results of the predictions of composite scores on pooled independent test sets of the IMAGEN, QTIM, QTAB and UKB cohorts.} We show the mean ICC and 95\% confidence intervals obtained through bootstrapping. Results are shown for
the Conv5-FC3 model trained using only IMAGEN (single) and IMAGEN, QTIM and QTAB (multi), alongside inter-rater and intra-rater performances. Three data augmentation strategies are compared (none, over-sampling (OS), flip). These results are shown for predictions in the left and right hemispheres.}
\end{figure}

    \begin{figure} [ht]
  \begin{center}
  \begin{tabular}{c} 
   \includegraphics[width=\textwidth]{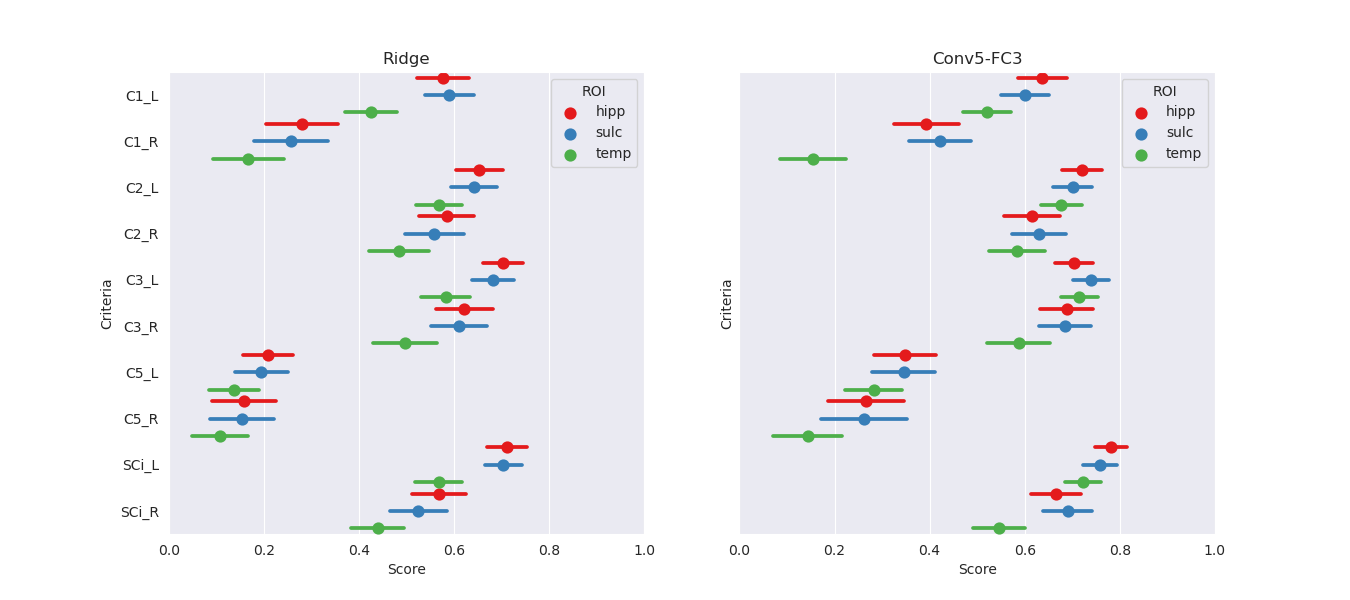}
  \vspace{-0.4cm}
  \end{tabular}
  \end{center}
  \caption[crit] 
  { \label{fig:ROI} 
  \emph{Perfomances using three ROIs.} We show the mean metrics (weighted kappas for C1 C2 and C3, an unweighted kappa for C5 and ICCs for SCi) and 95\% confidence intervals obtained through bootstraping. Results are shown for the ridge regression (left) and the Conv5-FC3 or CNN (right) for three ROIs: the hippocampus and close sulci (hipp), the hippocampus and all surrounding sulci (sulc) and the entire temporal lobe (temp).} 
  \end{figure}

\newpage
\subsection{Supplementary Methods and Results}
\subsubsection*{Methods}
\paragraph{ROI selection: }
Three ROIs were tested on a ridge regression and a Conv5-FC3 net in a previous study and evaluated on the independent test-set of the IMAGEN cohort. 
ROIs were as follows:
\begin{itemize}
    \item Hippocampus and close sulci: [24:96,54:107,16:49] in MNI coordinates
    \item Hippocampus and all surrounding sulci: [10:110,54:107,6:49] in MNI coordinates
    \item Temporal lobe: [10:110,15:107,6:79] in MNI coordinates
\end{itemize}

\paragraph{Data augmentation: }
As data, particularly in the right hemisphere, remains very unbalanced, we applied two data augmentation methods to improve our training. On one hand we simply over-sampled minority classes of each criterion by presenting the model with the same number of images from each class. This means that some images were shown repeatedly. Which images were shown several times was decided at random using sampling with replacement. 
On the other hand, we tried making up for the differences in scores in the left and in the right hemisphere by training not only on our original training set, but also on the same training set flipped vertically. In this way, the left hippocampus was found in the spot of the right hippocampus and vice-versa. The criteria were adjusted accordingly.

\subsubsection{Results}
\paragraph{ROI:} The effect of the ROI choice on the performance is presented on Figure~\ref{fig:ROI}. The smallest ROI achieved at least similar (if not better) performance on all criteria as larger ROI. As it is computationally more efficient to use a smaller ROI, we perform subsequent tasks using this ROI.
\paragraph{Data augmentation: }
We examine the performances of composite score predictions with over-sampling, with the addition of a flipped data-set and without data-augmentation on a pooled test set of all cohorts (N=502+248+100+246), comparing the results of the single training strategy (IMAGEN) and MUTLI (IMAGEN,QTIM,QTAB). Results are displayed in Figure~\ref{fig:CompositePoolAug} for the Conv-FC3 model. Human performances (inter and intra-rater ICCs) are plotted for reference.

None of our attempts at improving the training showed a significant improvement over the conv5-FC3 model trained on IMAGEN, QTIM and QTAB, in the right or left hemisphere. 

\end{document}